# Martensitic accommodation strain
## and the metal-insulator transition in manganites.

V. Podzorov[1], B. G. Kim[1], V. Kiryukhin[1], M. E. Gershenson[1] and S-W. Cheong[1,2]

[1] Rutgers University, Department of Physics & Astronomy, Piscataway, NJ 08854

[2] Bell Laboratories, Lucent Technologies, Murray Hill, NJ 07974

**ABSTRACT.**

In this paper, we report polarized optical microscopy and electrical transport studies of manganese oxides that reveal that the charge ordering transition in these compounds exhibits typical signatures of a *martensitic transformation.* We demonstrate that specific electronic properties of charge-ordered manganites stem from a combination of *martensitic accommodation strain* and effects of strong electron correlations. This intrinsic strain is strongly affected by the grain boundaries in ceramic samples. Consistently, our studies show a remarkable enhancement of low field magnetoresistance and the grain size effect on the resistivity in polycrystalline samples and suggest that the transport properties of this class of manganites are governed by the charge-disordered insulating phase stabilized at low temperature by virtue of martensitic accommodation strain. High sensitivity of this phase to strains and magnetic field leads to a variety of striking phenomena, such as unusually high magnetoresistance ($10^{10}$ %) in low magnetic fields.

Martensitic transformations, i.e. cooperative motion of atoms resulting in a formation of different crystal structure (martensitic phase or *martensite*) within a *parent* crystal, have been known in metals and alloys for more than a century[1-3]. In transition metal oxides, where strong electron-electron and electron-lattice interactions govern such phenomena as magnetic ordering, metal-insulator transition (MIT) and superconductivity, phase transitions are often accompanied by structural deformations[4]. However, the structural transformations are often considered a secondary or even a cumbersome effect. Here we demonstrate that the synergy between martensitic accommodation strain and strongly correlated electrons underlies the extraordinary electronic properties of the manganite compounds.

Recent research on manganites has revealed that the physical properties of these compounds are governed by coexistence and competition of different magnetic and structural phases - paramagnetic insulating, charge-ordered (CO) insulating and ferromagnetic metallic (FM) phases to list a few[4]. In particular, the coexistence of metallic and insulating phases leads to the percolative metal-insulator transition (MIT)[5-9], which is found to be very sensitive to a variety of parameters[10-13].

Our studies show that many of the transport properties of charge-ordered manganites are determined by the *martensitic nature* of the CO phase manifested by nucleation and growth of the lenticular shaped CO domains, the rapid propagation of domain walls through the crystal during the formation of martensite, and by thermal hysteresis of the resistivity ($\rho$) at the CO transition temperature ($T_{CO}$). Physical properties of martensitic alloys are governed by the long-range deformations of the crystal lattice, the so-called *accommodation strain*, which is induced as a result of nucleation of the martensitic particles within the parent matrix[1-3]. Growth of the accommodation strain with lowering the temperature (*T*) dominates the establishment of the thermo-elastic equilibrium between the parent phase and the martensite. This strain, inherent to martensites and transformation twins, is intrinsic and is different from the external strains associated with grain boundaries or substrate lattice mismatch in thin films[2].

In martensitic transformations, growth of martensitic domains across a grain boundary is prohibited. As a result, morphology of the martensitic particles and the accommodation strain strongly depend on the grain size. Concordantly, our studies of polycrystalline manganite samples demonstrate that $\rho$ becomes extremely sensitive to the grain size in the presence of the martensitic CO phase. The mechanism of this effect is essentially different from the spin-polarized tunneling across a grain boundary, which was found to be responsible for grain-boundary effects in non-charge-ordered manganite films[14-16].

The morphology of the CO phase and the transport properties of $Bi_{0.2}Ca_{0.8}MnO_3$, $Pr_{1/2}Ca_{1/2}MnO_3$ and $La_{0.275}Pr_{0.35}Ca_{0.375}MnO_3$ have been investigated by polarized optical microscopy at low temperatures. These compounds undergo a transition from the *parent* high-*T* cubic paramagnetic state to the *martensitic* CO insulating state at $T_{CO}$ = 180 K, 230 K and 210 K respectively [17, 11, 5]. Due to optical anisotropy of structurally distorted phases, the CO and cubic phases appear as bright and dark regions, respectively, in observation with $90^0$-crossed polarizer and analyzer, with an overall contrast increased when the distortion of the CO phase is stronger.

The CO transition in these compounds exhibits essential signatures of a martensitic transformation. For example, Fig. 1 shows a nucleation and growth of the CO domains within the parent cubic crystal in $Bi_{1-x}Ca_xMnO_3$ as a function of time, when $T$ is set and stabilized just below $T_{CO}$ = 180 K. Some of the domain walls were isothermally propagating through the crystal in a rapidly jerking motion - the phenomenon known in martensitic transformations as "umklapp" transformation[1,2]. All mentioned systems exhibit large accommodation strain at the CO domain boundaries. Indeed, the peculiar shape of the CO domains, thin plates or lenses, is a morphological signature of martensites and is dictated by minimization of the elastic energy of the structural distortion[1-3]. Transport measurements of single crystals of these compounds demonstrate a thermal hysteresis at $T_{CO}$, which is very typical for martensitic transformations and is a result of partial irreversibility of domain wall motion on cooling and heating. We note that the properties of other charge-ordered manganites are consistent with martensitic phenomenology. For example the CO transitions in $(Nd,Sm)_{1-x}Sr_xMnO_3$ (x≈1/2)[18] and in $La_{0.5}Ca_{0.5}MnO_3$[19] demonstrate signatures of martensitic transformations.

Grain boundaries are known to affect nucleation and growth of martensitic phases and, as a result, accommodation strain is expected to be very sensitive to the grain size in polycrystalline samples. In order to investigate the influence of grain boundaries on the accommodation strain, and therefore on physical properties of manganites, we have performed systematic studies of single crystalline and polycrystalline samples of the same chemical composition with different average grain size $<d>$ = 3, 6, 9, 12 and 17 µm. All the polycrystalline samples were prepared from a high-quality $La_{0.275}Pr_{0.35}Ca_{0.375}MnO_3$ pellet, which was synthesized by the standard solid-state reaction. The pellet was carefully ground to obtain fine powder and then a set of additional sintering at 1380 $^0$C was performed for different time periods $\Delta t$. The grain growth is described by the diffusion relationship $<d>^2 - d_0{}^2 = D\Delta t$, where $d_0$ and $D$ are constants (Fig. 2).

Phase morphology of individual grains in the polycrystalline La$_{0.275}$Pr$_{0.35}$Ca$_{0.375}$MnO$_3$ samples observed under a polarized optical microscope throughout the temperature range $T \approx 20 - 300$ K can also be understood in terms of large accommodation strain (Fig. 3). Above the CO transition, the crystal lattice of each grain is nearly cubic and, the micrograph at 300 K shows almost uniform dark surface. At $T$ near $T_{CO} \approx 210$ K, we observe nucleation and growth of small lens- and plate-shaped CO domains within each grain. With further cooling below $T_{CO}$, the dominant effect is an increase of the contrast of the image without further growth or shrinkage of the CO phase. The increase of the contrast is a result of increasing long-range structural deformation of the CO phase. The amount of the CO phase does not noticeably change even when the system is driven below the MIT by cooling (for this sample $T_{MI} \approx 130$ K, see $\rho(T)$ data below). This observation is consistent with recent x-ray and neutron scattering studies, reporting no reduction of the volume fraction of the CO below $T_{MI}$ in related samples[20-22]. Grain boundaries obviously influence the size and shapes of the CO domains. Nevertheless, we did not observe any simple morphological relationship between the grain boundaries and the location of nucleation sites of the CO domains, such as a preferential nucleation along the grain boundaries or, contrariwise, in the core of each grain.

Remarkable dependence of the transport properties on grain size $<d>$ is revealed by the resistivity ($\rho$) and low-field magnetoresistance ($MR \equiv (\rho_{H=0} - \rho_{5kOe})/\rho_{5kOe}$) measurements of polycrystalline La$_{0.275}$Pr$_{0.35}$Ca$_{0.375}$MnO$_3$ samples (Fig. 4). Data for a single crystal of the same composition are shown for comparison. Although $T_{CO} \approx 210$ K is similar for all samples, $T_{MI}$, defined as the temperature of the maximum of $d(log\rho)/dT$ on cooling, systematically decreases from 125 K to 30 K when $<d>$ is reduced from 17 to 6 μm, (the inset in Fig. 4). Finally, the specimen with the smallest grain size, 3 μm, does not exhibit the MIT down to 20 K, below which $\rho$ becomes too large to be measured reliably. Surprisingly, a change of the grain size by a factor of two is sufficient to switch the low-$T$ state of the same compound from metallic to insulating. The low-$T$ resistivity, $\rho_0 \equiv \rho(20$ $K)$, appears to be the most grain-size-sensitive characteristic of the samples. For $<d>$ = 3 - 17 μm, $\rho_0$ varies systematically over the range $0.1 - 10^8$ Ωcm in $H = 0$.

Sensitivity of $\rho_0$ and $T_{MI}$ to the grain size $<d>$ becomes much less pronounced when a small magnetic field $H = 5$ kOe is applied (Fig. 4, the lower panel). Firstly, this field induces the MIT in the sample with $<d>$ = 3 μm. Secondly, $\rho(T)$ curves for all samples become much more similar in $H = 5$ kOe, e.g. $\rho_0$ and $T_{MI}$ fall into much narrower ranges for all $<d>$. This behavior rejects possible arguments that the strong dependence of $\rho_0$ on $<d>$ shown in the upper panel is due to the morphological defects of ceramic samples such as grain boundaries or pores, as these "permanent" defects would not be sensitive to the magnetic field. Low-field $MR_{20K}$ in $H = 5$ kOe varies systematically with $<d>$ = 3 - 17 μm over a broad range $0.1 - 10^8$ (see the inset). Such a large low-field $MR$ cannot be explained by the spin-polarized transport across magnetic domain boundaries, which has been shown to result in $MR$ = 20-30 % in systematically studied epitaxial films grown on bicrystal substrates and in polycrystalline films[14-16].

It is well known that the magnetic field of 5 kOe is not strong enough to "melt" the CO phase and results only in rotation of the existing FM moments in the sample[11, 22].

However, the strong suppression of the grain-size dependence of $\rho_0$ by $H$=5 kOe is obvious after a comparison of the upper and the lower panels of Fig. 4. This suggests that the insulating phase responsible for the large $\rho_0$ in $H$ = 0 is much more sensitive to the magnetic field than CO and, therefore, cannot be the CO phase. This result is consistent with our optical studies (Fig. 3), where no changes of the volume fraction of the CO phase were observed when the sample is driven through the MIT. Therefore, both our transport and optical experiments indicate that the martensitic CO phase of this system does not undergo the MIT and can only affect the properties of other phases, which undergo the transition.

On the basis of our optical observations and transport measurements, we propose the following scenario for the MIT in these CO manganese oxides. The accommodation strain, introduced by the martensitic CO domains into the surrounding lattice at $T_{CO} \approx 210$ K, significantly affects the properties of the parent paramagnetic phase of $La_{0.275}Pr_{0.35}Ca_{0.375}MnO_3$. When this phase is loaded with the strain, its FM transition is suppressed, e.g. $T_{MI}$ is shifted to a lower temperature. Therefore, the parent phase tends to retain its high-$T$ paramagnetic properties, remaining *charge-disordered* and *insulating* (CD-I) even at low $T$. It is known, that the martensitic strain is more difficult to accommodate as the grain size of polycrystalline samples decreases [2, 3]. As a result, the amount of strain-loaded CD-I phase is higher in the samples with smaller grain size. The presence of this insulating phase results in an unusually high resistivity below the MIT and leads to the low temperature shift of the transition with decreasing $<d>$. Our optical observations show that the CO phase is not involved in the MIT directly. Thus, the transition occurs within the parent phase, when it separates into the FM metallic and strain-stabilized CD-I phases at $T \leq T_{MI}$. It is noteworthy that the ability of strain to stabilize phases, which do not exist at all without the strain, is well known in conventional martensites[1]. For an illustrative comparison with a system without intrinsic martensitic strain, recall $La_{1-x}Ca_xMnO_3$ ($0.2<x<0.5$) where there is no CO and, hence, little or no internal strain is present. The MIT in this system occurs at much higher temperature ($T_{MI} \sim 275$ K) and resembles a second order phase transition[23].

In summary, we found that the martensitic accommodation strain dominates the physics of the charge-ordering transformation in manganites. Sensitivity of the martensitic phase to the grain boundaries leads to the observed striking dependence of the transport properties of polycrystalline samples on grain size. Our optical microscopy and magneto-transport measurements at low temperature indicate that the *charge-disordered insulating* phase, which is responsible for the metal-insulator transition in the charge-ordered manganites, is the parent paramagnetic phase stabilized at low temperature by the martensitic accommodation strain. Various signatures of martensitic transformations manifested in mixed-valent manganites indicate general applicability of the martensitic approach and phenomenology to the structural phase transitions in oxides with strongly correlated electrons.

**Acknowledgments.**

This work was supported by the NSF under Grant No. DMR-9802513 and by the MRSEC program of the NSF, Grant No. DMR-0080008.

**Figure Captions**

**1.** Micrographs of $Bi_{0.2}Ca_{0.8}MnO_3$ single crystal taken with a polarized optical microscope just below $T_{CO} = 180$ K in time intervals $\Delta t = 1$s. Bright regions correspond to the CO domains; dark ones to the cubic parent lattice. Nucleation and growth of the plate and lenticular shaped CO domains is evident.

**2.** Micrographs of polished polycrystalline $La_{0.275}Pr_{0.35}Ca_{0.375}MnO_3$ samples sintered at $T$ = 1380 $^0$C for time intervals $\Delta t = 0$ (initial pellet), 10, 40 and 100 hours (panels a, b, c and d, respectively). The average grain size $<d>$ is indicated. The dependence $<d>^2$ vs. $\Delta t$ is shown in the lower panel.

**3.** Polarized optical micrographs of the polycrystalline $La_{0.275}Pr_{0.35}Ca_{0.375}MnO_3$ sample with average grain size $<d> = 17$ μm and the MIT temperature $T_{MI} \approx 130$ K, taken at various temperatures on cooling. Bright regions correspond to the CO phase; dark ones – to the cubic parental lattice.

**4.** Temperature dependence of the resistivity $\rho(T)$ in zero magnetic field $H = 0$ (upper panel) and in 5 kOe (lower panel) of the polycrystalline $La_{0.275}Pr_{0.35}Ca_{0.375}MnO_3$ samples with different grain size $<d> = 3$ - $17$ μm and a single crystal of the same composition measured during cooling and heating. The upper inset shows the dependence of the MIT temperature, $T_{MI}$, on $<d>$ for zero field cooling (closed circles) and for field cooling in 5 kOe (open circles). The lower inset shows the grain-size dependence of the magnetoresistance $MR$ in 5 kOe at $T = 20$ K.

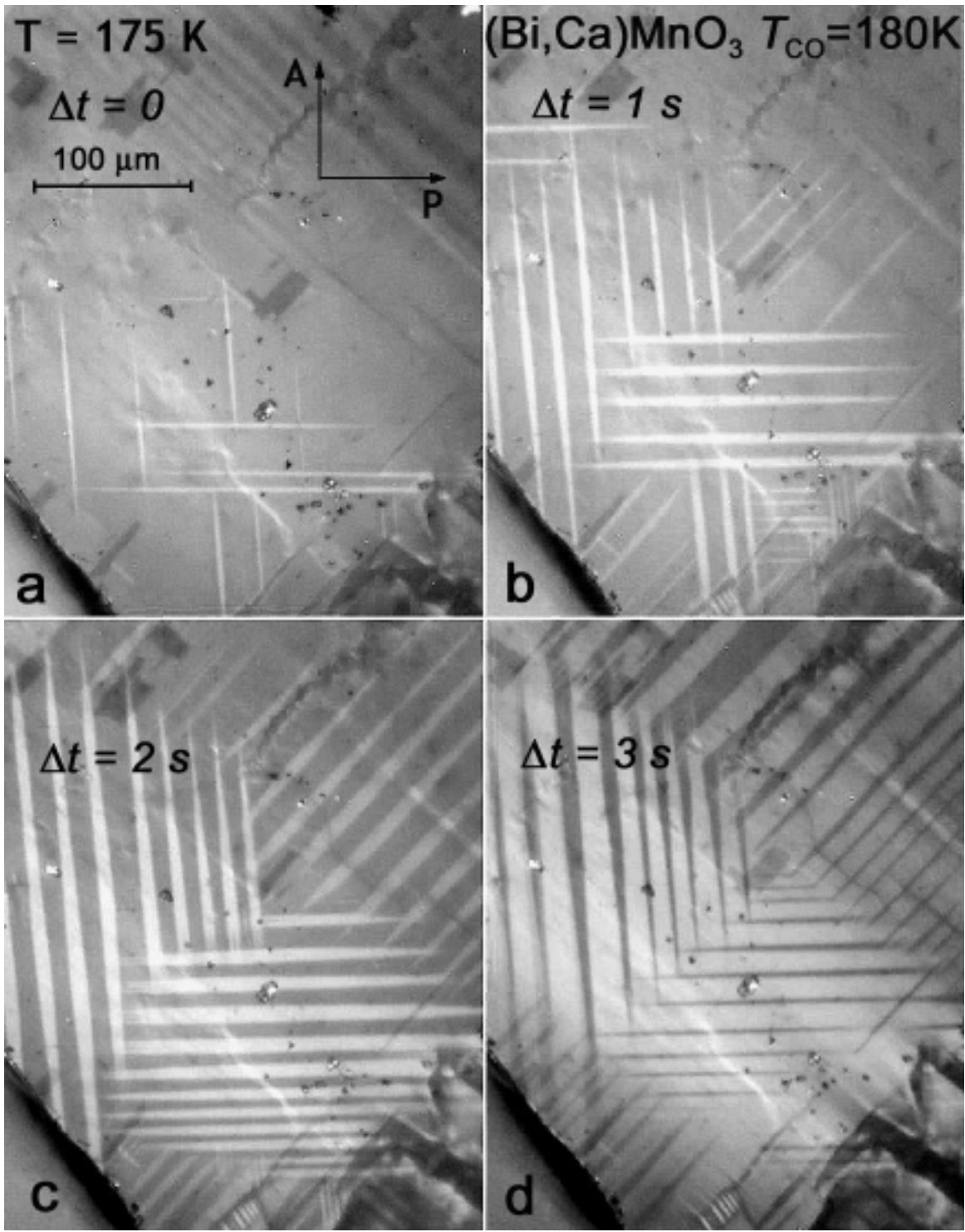

T = 175 K    (Bi,Ca)MnO₃ $T_{CO}$=180K

$\Delta t = 0$    $\Delta t = 1\ s$

100 μm

A

P

a    b

$\Delta t = 2\ s$    $\Delta t = 3\ s$

c    d

Fig. 1

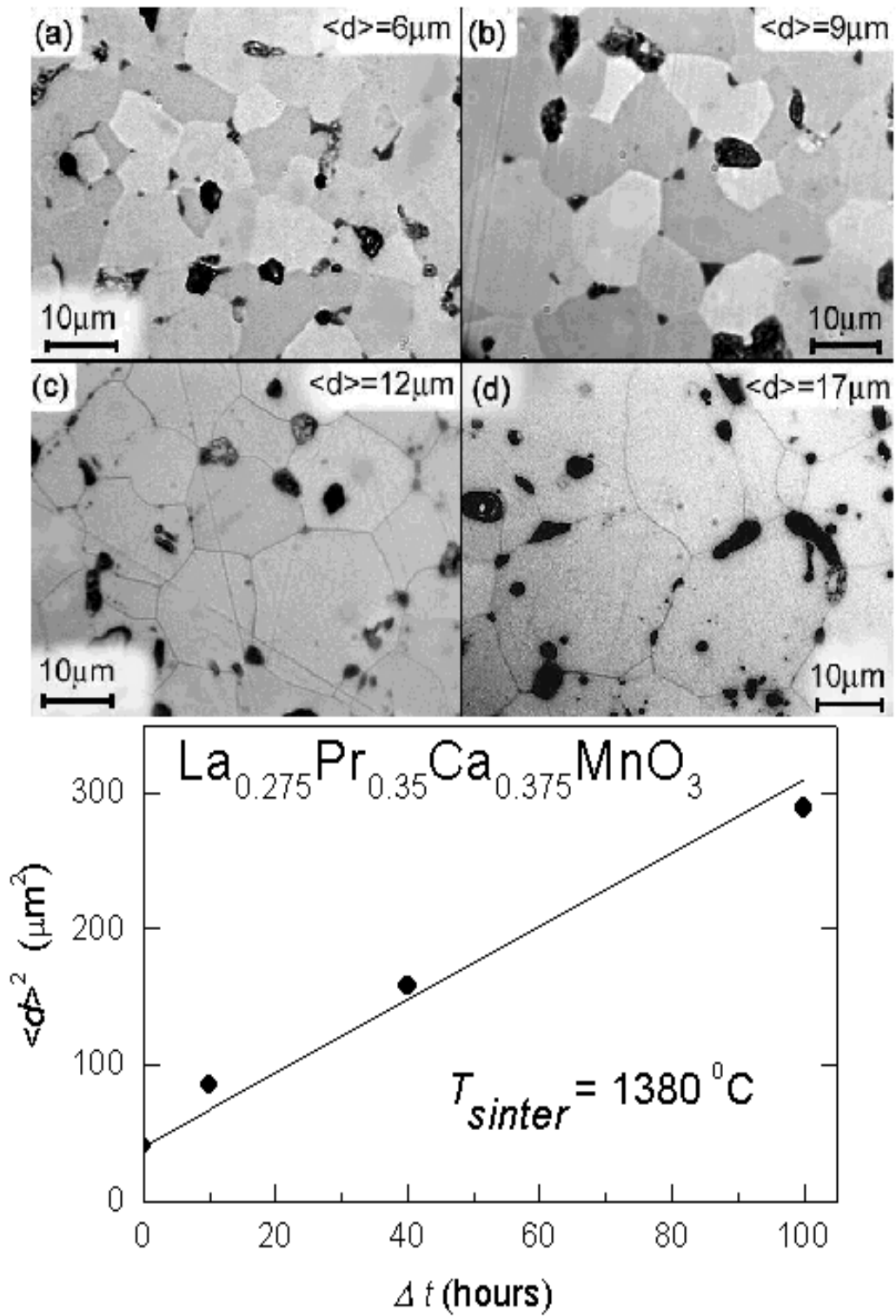

Fig. 2

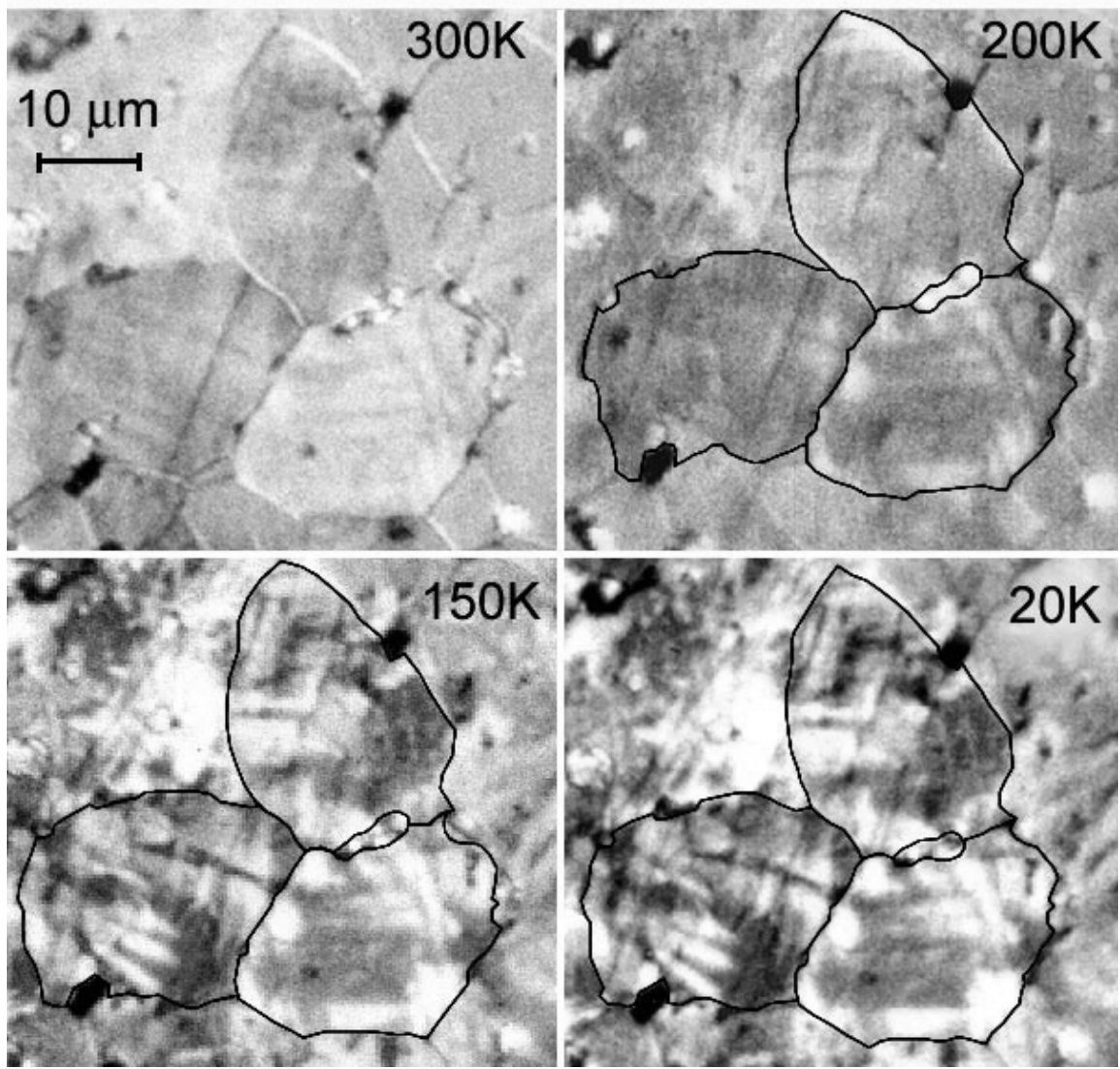

Fig. 3

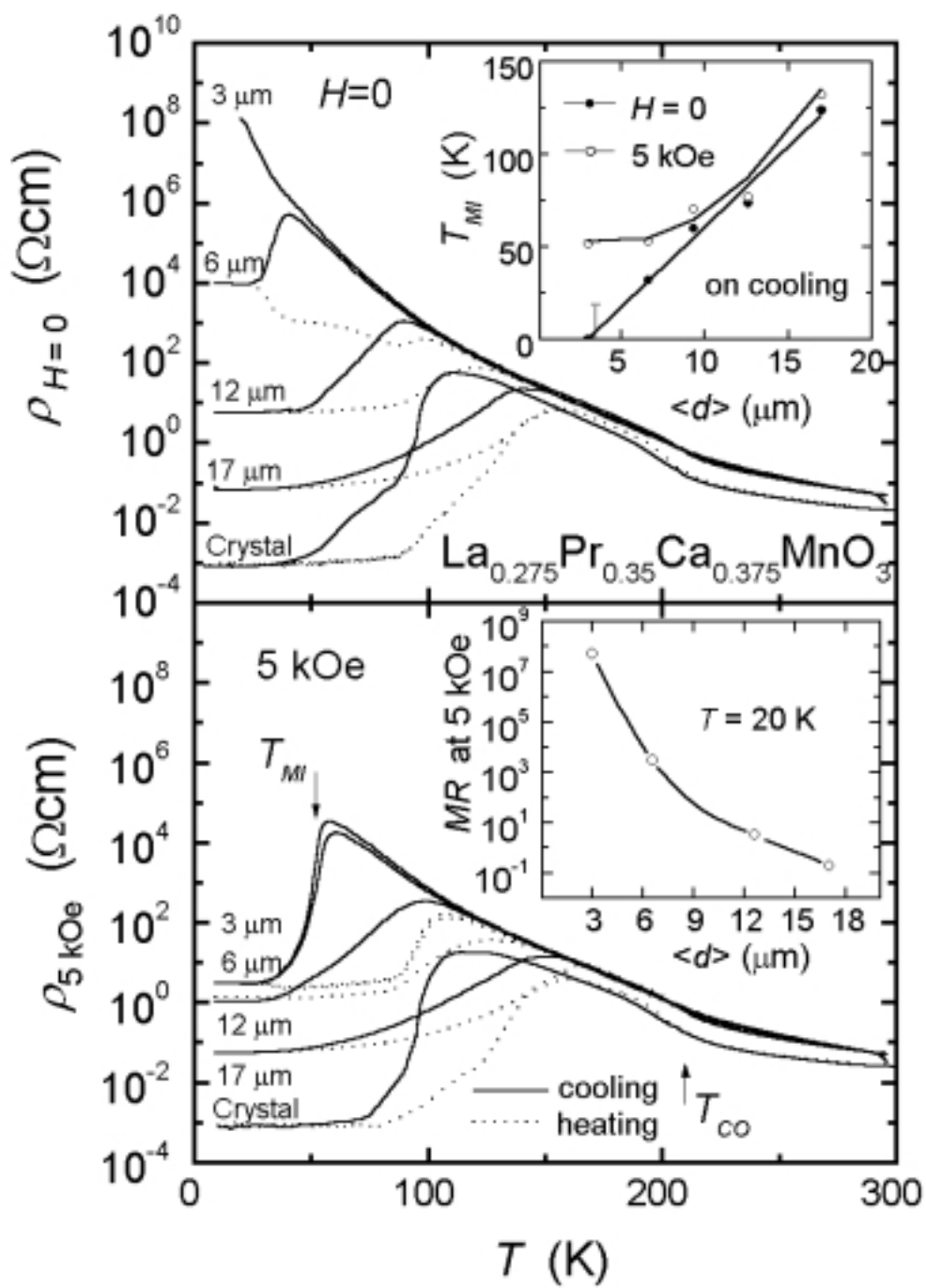

Fig. 4